\documentclass[aps,prd,twocolumn,showpacs,amsmath,amssymb,superscriptaddress,nofootinbib,floatfix,preprintnumbers,secnumarabic]{revtex4-1}
\pdfoutput=1

\usepackage[usenames,dvipsnames]{color}
\usepackage{multirow}
\usepackage{graphicx}
\usepackage{adjustbox}
\usepackage{booktabs}
\usepackage{appendix}
\usepackage{slashed}
\usepackage{url}
\usepackage{xspace}
\usepackage{braket}
\usepackage{tabularx}
\usepackage[normalem]{ulem} 
\usepackage{siunitx}
\usepackage[ISO]{diffcoeff}
\usepackage[utf8]{inputenc}
\usepackage{xspace}
\usepackage{bm}
\definecolor{darkblue}{rgb}{0,0,0.5}
\definecolor{darkgreen}{rgb}{0.0,0.5,0.2}
\definecolor{darkred}{rgb}{0.6,0,0}
\usepackage[pdfstartview=XYZ,
bookmarks=true,
colorlinks=true,
linkcolor=darkred,
urlcolor=blue,
citecolor=darkgreen,
pdftex,
bookmarks=true,
breaklinks=true,
linktocpage=true, 
hyperindex=true
]{hyperref}
\usepackage[capitalise]{cleveref}
\usepackage{orcidlink}

\usepackage{fullpage}
\usepackage{geometry}      
\AtBeginDocument{
  \heavyrulewidth=.08em
  \lightrulewidth=.05em
  \cmidrulewidth=.03em
  \belowrulesep=.65ex
  \belowbottomsep=0pt
  \aboverulesep=.4ex
  \abovetopsep=0pt
  \cmidrulesep=\doublerulesep
  \cmidrulekern=.5em
  \defaultaddspace= .5em
  \setlength{\tabcolsep}{0.5em}
}

\begin{document}

\title{Cosmological impact of $\nu$DM interactions\\ enhanced in narrow redshift ranges}


\author{Sebastian Trojanowski}
\email{Sebastian.Trojanowski@ncbj.gov.pl}
\affiliation{National Centre for Nuclear Research, Pasteura 7, Warsaw, PL-02-093, Poland} 
\author{Lei Zu}
\email{Lei.Zu@ncbj.gov.pl}
\affiliation{National Centre for Nuclear Research, Pasteura 7, Warsaw, PL-02-093, Poland} 

\date{\today}

\begin{abstract}
The impact of dark matter-neutrino ($\nu$DM) interactions on cosmological perturbations has regained attention, spurred by indications of non-zero couplings from high-multipole cosmic microwave background data, weak lensing, and Lyman-$\alpha$ observations. We demonstrate that a similar observational preference is obtained if $\nu$DM interactions are primarily enhanced during a specific epoch, $z\sim (10^4-10^5)$, leading to $>3\sigma$ preference for a non-zero interaction in the combined Atacama Cosmology Telescope and cosmic shear data. This redshift-limited enhancement circumvents other cosmological and astrophysical bounds and can be achieved within a neutrino portal dark matter framework incorporating resonantly enhanced scattering rates.
\end{abstract}

\maketitle

\section{Introduction\label{sec:intro}}

From its earliest stages, the evolution of the universe was driven by the interplay between radiation and matter. Relic neutrinos and dark matter (DM) played a crucial role in this process before recombination, with DM continuing to shape matter distribution in later epochs. Potential interactions between these two components could significantly affect the predictions of the standard cosmological model ($\Lambda$CDM)~\cite{Boehm:2000gq,Boehm:2004th,Mangano:2006mp,Serra:2009uu,Shoemaker:2013tda,Wilkinson:2014ksa,Escudero:2015yka,Stadler:2019dii,Mosbech:2020ahp,Paul:2021ewd,Pal:2023dcs}.

Recent analyses of Lyman-$\alpha$ data~\cite{Hooper:2021rjc}, cosmic microwave background (CMB) measurements~\cite{Brax:2023rrf,Brax:2023tvn,Giare:2023qqn}, and weak lensing (WL) cosmic shear observations~\cite{Zu:2025lrk} have suggested possible non-negligible DM interactions with neutrinos ($\nu$DM). The latter two hints point to a characteristic interaction strength, parameterized by $u_{\nu\textrm{DM}}$, defined as:
\begin{equation}
u_{\nu\textrm{DM}} = \frac{\sigma_{\nu \rm{DM}}}{\sigma_{T}}\left(\frac{m_\chi}{100~\textrm{GeV}}\right)^{-1} \sim 10^{-4},
\label{eq:unuDM}
\end{equation}
where $m_\chi$ is the DM mass, and $\sigma_{\nu \rm{DM}}$ and $\sigma_T$ are the $\nu$DM and Thomson scattering cross sections, respectively. However, strong constraints on $\sigma_{\nu \rm{DM}}$ from astrophysics~\cite{Farzan:2014gza,Arguelles:2017atb,Pandey:2018wvh,Kelly:2018tyg,Alvey:2019jzx,Choi:2019ixb,Jho:2021rmn,Ghosh:2021vkt,Lin:2022dbl,Cline:2022qld,Ferrer:2022kei,Cline:2023tkp,Lin:2023nsm,Fujiwara:2023lsv,Heston:2024ljf,Lin:2024vzy,Fujiwara:2024qos,Zapata:2025huq,Leal:2025eou,Chauhan:2025hoz} and small-scale structure observations~\cite{Boehm:2014vja,Dey:2022ini,Akita:2023yga,Dey:2023sxx,Crumrine:2024sdn} challenge such large values of $u_{\nu \rm{DM}}$. This is particularly true for a constant thermally-averaged $\nu$DM cross section, $\sigma_{\nu\textrm{DM}}\propto T_\nu^0$, or simple power-law dependencies, such as $\sigma_{\nu\textrm{DM}}\propto T_\nu^2$, which are predicted in many effective scenarios~\cite{Olivares-DelCampo:2017feq}.

In realistic $\nu$DM models, though, the interaction cross section is expected to exhibit a more complex temperature dependence. Even if $\sigma_{\nu\textrm{DM}}\propto T_\nu^n$ (where $n = 2,4,\ldots$) at late times, unitarity considerations necessitate a breakdown of this dependence at higher temperatures, preventing the cross section from growing arbitrarily large in the early universe. This can naturally lead to a maximum $\nu$DM interaction strength within a specific temperature range.

Resonantly enhanced $\sigma_{\nu\textrm{DM}}$ is a particularly intriguing possibility, potentially increasing the interaction rate by orders of magnitude within a narrow energy window. As the universe cooled, the neutrino temperature and characteristic energy decreased from $T_\nu\sim 1~\textrm{MeV}$ at weak interaction decoupling (redshift $z_{\textrm{weak}}\sim 10^{10}$) to $T_\nu\lesssim \mathcal{O}(10~\textrm{meV})$ at the neutrino non-relativistic transition ($z_{\textrm{non-rel.}} \sim 100$). If resonant $\nu$DM interactions occur within this energy range, the enhanced cross section within a specific energy window would inevitably imprint distinctive features on the corresponding scale structure. 

\begin{figure*}[t!]
\centering
\includegraphics[scale=0.9]{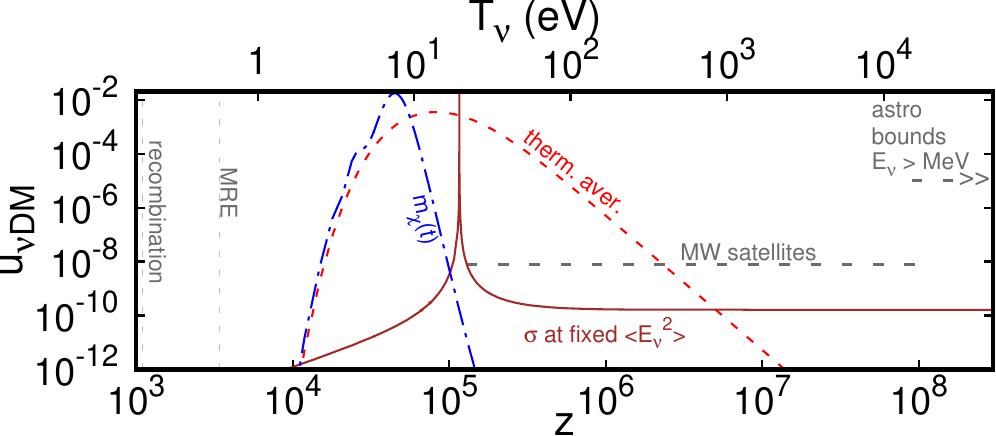}
\caption{The dependence of the $\nu$DM interaction parameter, $u_{\nu\textrm{DM}}$, on redshift $z$ (bottom axis) and neutrino temperature $T_\nu$ (top axis). The brown solid line is derived assuming a fixed neutrino energy given by its thermal average, $E_\nu^2 = \langle E_\nu^2\rangle$. The red dashed line represents a more realistic calculation involving a convolution with the neutrino energy spectrum. The blue dot-dashed line combines this convolution with a mild time-dependence of the DM mass. The results correspond to the neutrino portal model with a resonantly enhanced $\nu$DM cross section; see the text for details. Approximate bounds from Milky Way satellite galaxies for $z\gtrsim \textrm{a few}\times 10^5$ are indicated. Astrophysical constraints further limit the allowed $\nu$DM cross sections at high energies ($E_\nu> \textrm{MeV}$), as schematically depicted in the plot.}
\label{fig:resonantunuDM}
\end{figure*}

In this paper, we investigate the cosmological impact of such enhanced $\nu$DM interactions on CMB and cosmic shear observations, where hints of non-zero $u_{\nu \textrm{DM}}$ have been reported. We demonstrate that a narrow but significant enhancement of $\sigma_{\nu\textrm{DM}}$ can reconcile these observations with existing constraints, particularly if the enhancement occurs for redshifts of $z\sim 10^4-10^5$. This is schmatically illustrated in \cref{fig:resonantunuDM}, which displays selected constraints and their corresponding neutrino temperatures and redshifts, alongside an example of a resonant $u_{\nu \textrm{DM}}$ temperature dependence motivated by a benchmark beyond the Standard Model (BSM) scenario discussed below.

The paper is organized as follows. \Cref{sec:data} identifies cosmologically viable redshift ranges and values of $u_{\nu\textrm{DM}}$ by examining a toy $\nu$DM model, in which $\sigma_{\nu\textrm{DM}}\propto T_\nu^0$ is assumed within a limited redshift range. \Cref{sec:bsm} explores sample benchmark scenarios within a specific neutrino portal DM model with resonantly enhanced $\sigma_{\nu\textrm{DM}}$. \Cref{sec:conclusions} summarizes our findings. \Cref{app:simulation} details the cosmological data analysis.

\section{$\nu\textrm{DM}$ at specific redshifts\label{sec:data}}

The impact of $\nu$DM interactions on perturbation evolution is characterized by additional terms in the Boltzmann hierarchy, which modify the energy and momentum transfer between DM particles and neutrinos~\cite{Ma:1995ey,Cyr-Racine:2015ihg}. For cold dark matter (CDM), the evolution of the fluid velocity divergence $\theta_{\chi}$ includes a contribution proportional to the $\nu$DM interaction rate, $\dot{\mu}_\chi$, as given by:
\begin{equation}
\dot{\theta}\chi = k^2\psi - \mathcal{H}\theta\chi - K_\chi\,\dot{\mu}\chi\,(\theta\chi-\theta_\nu),
\label{eq:thetachi}
\end{equation}
where $\phi$ and $\psi$ are scalar potentials describing metric perturbations, and $\mathcal{H} = \dot{a}/a$ represents the Hubble rate. For massless neutrinos, $K_\chi = (4/3)\rho_\nu/\rho_\chi$ and $\dot{\mu}_\chi = a\,n_\chi\,\sigma_{\nu\textrm{DM}}$, assuming a constant cross section, where $n_\chi = \rho_\chi/m_\chi$ is the CDM number density for a DM particle mass of $m_\chi$. The Boltzmann hierarchy for neutrinos is similarly modified, and the relevant interaction terms can be effectively represented by the $u_{\nu \textrm{DM}}$ parameter, as defined in \cref{eq:unuDM}~\cite{Mangano:2006mp,Wilkinson:2014ksa,Mosbech:2020ahp}.

In the redshift range $10^2 \ll z \ll 10^{10}$, neutrinos remain relativistic, but their energy is significantly lower than the DM mass, $E_\nu\ll m_\chi$, assuming $m_\chi\gtrsim 10~\textrm{MeV}$. Consequently, $\nu$DM interactions are conventionally modeled as Thomson-like scattering, analogous to electron-photon interactions prior to recombination~\cite{Dodelson:1993xz}. We begin with applying this common approximation to identify viable cosmological redshift ranges. Instead of considering a strict temperature-independent cross section, we assume that $\sigma_{\nu\textrm{DM}}$ is non-zero and constant only within a limited redshift range. In addition to the standard six $\Lambda$CDM parameters, this analysis introduces three new parameters: the interaction strength, $u_{\nu\textrm{DM}}$, the initial redshift at which this parameter becomes non-zero, $z_{\textrm{min}}$, and the length of the redshift range, $\Delta z$. The strength of the $\nu$DM interaction is assumed to be negligible outside the interval $[z_{\textrm{min}},z_{\textrm{min}} + \Delta z]$.

We tested this scenario against the Dark Energy Survey (DES) three-year cosmic shear data~\cite{DES:2021bvc}, the CMB likelihoods from the Planck 2018 Legacy~\cite{Planck:2019nip}, the Atacama Cosmology Telescope (ACT) DR4 dataset, and the Baryon Acoustic Oscillation (BAO) data from the BOSS DR16 data release~\cite{eBOSS:2020tmo,eBOSS:2020hur,eBOSS:2020uxp,eBOSS:2020fvk,eBOSS:2020yzd}. We employ a modified version of the \texttt{CLASS} code~\cite{Stadler:2019dii,Mosbech:2020ahp}. When analyzing WL data, we include non-linear corrections to the matter power spectrum by interpolating between the results of $N$-body simulations~\cite{Zhang:2024mmg,Zu:2025lrk}. Details of the analysis are given in \textsl{Supplementary Material}.

\Cref{fig:simresults} displays the resulting posterior distributions. The left panel shows the 1D posterior for the neutrino interaction parameter, contrasting our result with that for a redshift-independent cross section denoted as ``$u_{\nu\textrm{DM}} = \textrm{const}$''~\cite{Zu:2025lrk}. Our analysis favors larger interaction strengths due to their redshift-limited nature, shifting the posterior for $u_{\nu\textrm{DM}}$ to higher values. This results in $>3\sigma$ preference for a non-zero interaction, peaking at $u_{\nu\textrm{DM}}\sim 10^{-2.5}$ with a $95\%$ lower bound of $u_{\nu\textrm{DM}} > 10^{-3.65}$; see \cref{tab:benchmarks}.

This strong preference is driven by our core assumption that $\nu$DM interactions are localized in time. The right panel of \cref{fig:simresults} illustrates this with the posterior in the $(z_{\textrm{min}}, \Delta z)$ plane. The $1\sigma$ contour favors an interaction epoch starting at $z_{\textrm{min}}\sim 10^4-10^5$ and extending to $\Delta z\sim 10^5$, and potentially beyond. We note, however, that increasing the interaction duration, $\Delta z$, primarily affects small-scale perturbations that enter the horizon at earlier times. Our analysis deliberately masks these scales (at $k\gtrsim \textrm{a few}~h/\rm{Mpc}$) in the weak lensing data to avoid potential biases from baryonic feedback~\cite{DES:2021bvc}. The resulting insensitivity of our likelihood explains the extended tail of the $\Delta z$ posterior. Nevertheless, as discussed below, other small-scale probes like dwarf galaxy counts likely impose an upper limit of $\Delta z\lesssim 10^6$, which we adopted as a prior in our analysis.

\begin{figure*}[t!]
\centering

\raisebox{0.15cm}{\includegraphics[width=0.58\textwidth]{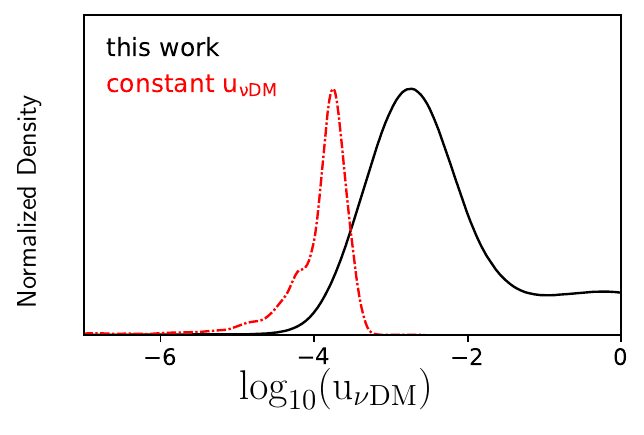}}
\includegraphics[width=0.40\textwidth]{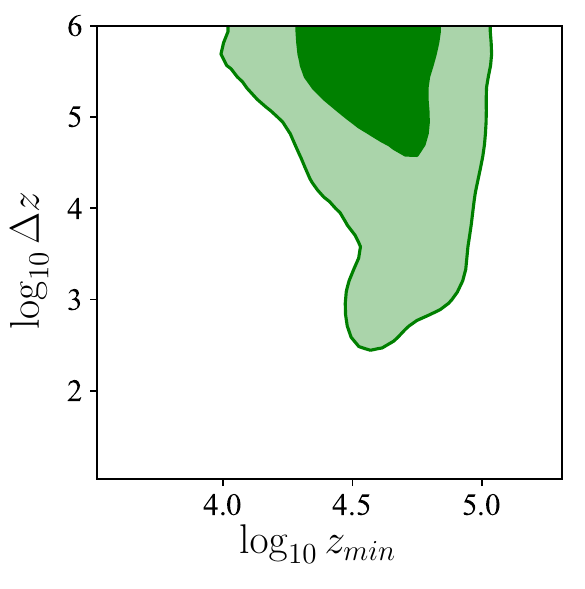}

\caption{\textsl{Left panel}: The one-dimensional posterior distributions for the parameter $\textrm{u}_{\nu\rm{DM}}$ in this work and the ``$u_{\nu\textrm{DM}} = \textrm{const}$'' scenario. \textsl{Right panel}: The marginalized 2D posterior distribution in the ($z_{min},\Delta z$) plane. The contours show the $68\%$ (inner contour) and $95\%$ (outer contour) credible regions.}
\label{fig:simresults}
\end{figure*}

\begin{table*}
    \centering
    \begin{tabular}{|c|c||c|c|c|c|c|}
\hline
\multirow{2}{*}{Parameter} & Credible & \multicolumn{4}{c|}{Benchmark models} & $u_{\nu\textrm{DM}} = \textrm{const}$ \\
\cline{3-6}
& regions & BP1  & BP2 & BP3 & BP4 & \scriptsize{(BFP)} \\
\hline
\hline
$100 \Omega_b h^2$ & $2.226_{-0.015}^{+0.015}$ & 
2.229 & 2.216  & 2.212 & 2.238
& 2.238\\
$\Omega_{m} $ & & 0.3090 &  
0.3064 & 0.3063 & 0.2997  & 0.3061 \\
$100 \theta_s$ & $1.04219_{-0.00031}^{+0.00050}$ & 1.042 &
1.043 & 1.042 & 1.042 & 1.043 \\
$\ln \left(10^{10} A_s\right)$ & $3.022_{-0.014}^{+0.017}$ & 3.031 &
3.014 & 3.023 & 3.037 & 3.039 \\
$n_s$ & $0.9634_{-0.0067}^{+0.0075}$ & 0.9643 &  
0.9634 & 0.9639 & 0.9723 & 0.9736 \\
$\tau_{\text {reio }}$ & $0.0470_{-0.0068}^{+0.0081}$ & 0.0499 & 
0.0437 & 0.0508 & 0.0532 & 0.0536 \\
\hline\noalign{\vspace{0.25ex}}
\cline{1-7}
$\log_{10}\rm{u}_{\nu\rm{DM}}$ & $ -2.46^{+0.49}_{-1.0}(>-3.65)$ & -2.910 & 
-2.127 & -1.575 & -0.8706 & -3.696 \\
$\log_{10} z_{\textrm{min}}$ & $4.58_{-0.14}^{+0.17}$ & 4.565 &
4.673 & 4.805 & 3.533 & - \\
$\log_{10}\Delta z$ & $5.25^{+0.72}_{-0.12}$ & 5.977 &  
4.754 & 4.189 & 1.494 & - \\
\cline{1-7}
\multicolumn{1}{c}{} & \multicolumn{1}{r|}{$\chi^2 - \chi^2_{\Lambda\textrm{CDM}}$} & -16.6 & 
-11.0 &-7.70& -5.40 & -7.06 \\
\multicolumn{1}{c}{} & \multicolumn{1}{r|}{$S_8$} & 0.8005 & 0.8055 & 0.8185 & 0.7989  & 0.7724 \\
\cline{3-7}
\end{tabular}
    \caption{The 68\% (95\%) credible regions and benchmark points from the MCMC simulation chains, including the six standard cosmological parameters along with $u_{\nu\rm{DM}}$, $z_{\textrm{min}}$, and $\Delta z$. The best-fit point for the ``$u_{\nu\textrm{DM}} = \textrm{const}$'' scenario is shown for comparison.} 
    \label{tab:benchmarks}
\end{table*}

\Cref{tab:benchmarks} additionally details several benchmark models from our analysis that explore a range of $\Delta z$ and $z_{\textrm{min}}$ values, alongside the best-fit results for the ``$u_{\nu\textrm{DM}} = \textrm{const}$'' scenario~\cite{Zu:2025lrk}. For each case, we report the improvement in fit relative to $\Lambda$CDM, $\chi^2-\chi^2_{\Lambda\textrm{CDM}}$. Specifically, our best-fit point (BP1) confines the $\nu$DM interaction to $3.7\times 10^4 \lesssim z \lesssim 10^6$, yielding an improved fit of $\chi^2-\chi^2_{\Lambda\textrm{CDM}} = -16.6$. Compared to the ``$u_{\nu\textrm{DM}} = \textrm{const}$'' model, our scenario has two additional free parameters, resulting in a $2.3\sigma$ preference for BP1. This is achieved with a large cross section, $u_{\nu\textrm{DM},\textrm{BP1}}\sim 10^{-3}$ within the aforementioned redshit range.

\begin{figure*}[t!]
\centering
\includegraphics[width=0.49\textwidth]{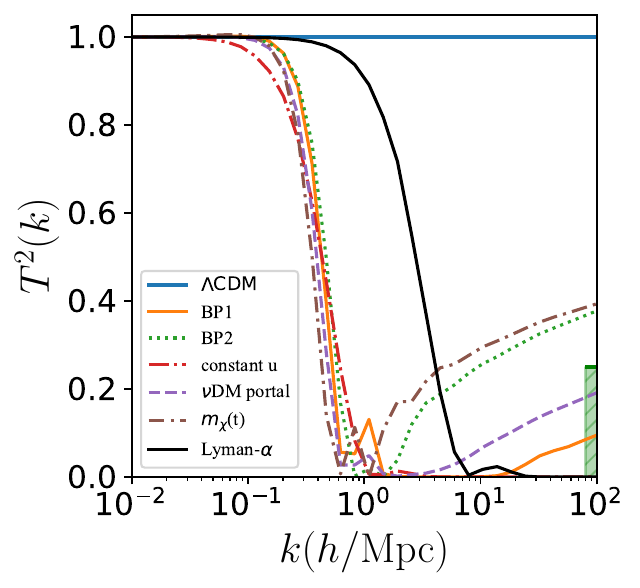}
{\includegraphics[width=0.465\textwidth]{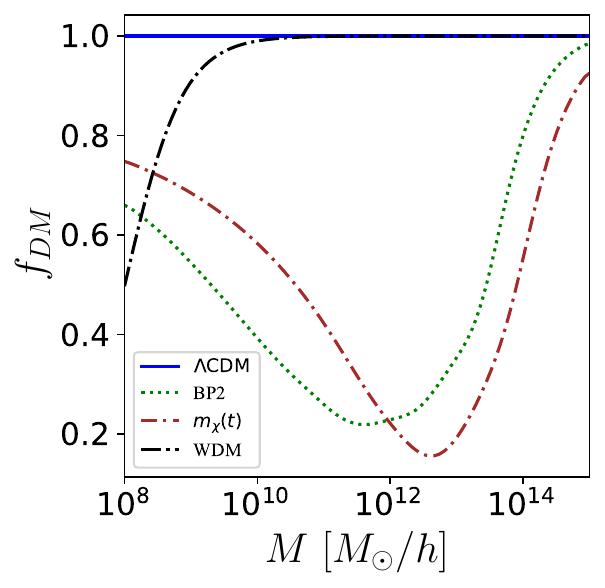}}
\caption{\textsl{Left panel}: The linear transfer function for BP1, BP2, and the ``$u_{\nu\textrm{DM}} = \textrm{const}$'' scenario (see \cref{tab:benchmarks}), along with the benchmark points for the neutrino-portal DM model without and with time-varying DM mass. The result for the best-fit point obtained in the Lyman-$\alpha$ analysis is shown for comparison~\cite{Hooper:2021rjc}. Constraints from dwarf galaxy counts are illustrated with the green-shaded region. \textsl{Right panel}: The ratio $f_{\textrm{DM}}$ of the subhalo mass function for selected $\nu$DM benchmark scenarios discussed in the text with respect to $\Lambda$CDM, as a function of the subhalo mass $M$. For comparison, the result for warm dark matter with $m_{\textrm{WDM}} = 6.5~\textrm{keV}$ is shown with a black dot-dashed line.}
\label{fig:halomassf}
\end{figure*}

The table also includes other benchmarks (BP2-4) with progressively smaller values of $\Delta z$. While these points have a weaker overall fit to CMB and WL data than BP1, their more localized interaction at $z\sim 10^4-10^5$ causes less suppression of the  matter power spectrum at $k\gtrsim 10~h/\rm{Mpc}$. This is illustrated in the left panel of \cref{fig:halomassf}, which shows the linear transfer function ratio, $T^2(k) = P_{\nu\textrm{DM}}(k)/P_{\Lambda\textrm{CDM}}$. Unlike BP1, which shows strong suppression at $k\gtrsim 10~h/\rm{Mpc}$, similar to the ``$u_{\nu\textrm{DM}} = \textrm{const}$'' case, the BP2 model exhibits only mild suppression, with $T^2(k)\sim 0.4$.

This reduced suppression is significant, as small-scale structure places stringent constraints on $\nu$DM interactions, potentially limiting the coupling to $u_{\nu\textrm{DM}} \lesssim 10^{-8}$~\cite{Crumrine:2024sdn}. This bound arises from adapting warm dark matter (WDM) constraints on Milky Way (MW) dwarf counts~\cite{DES:2020fxi}, which require that the $\nu$DM power suppression not exceed that of a $6.5~\textrm{keV}$ WDM particle. This translates to a lower limit on the transfer function, $T^2(k_{\textrm{hm}})\gtrsim 0.25$ at $k_{\textrm{hm}} \simeq 81~h/\rm{Mpc}$. The corresponding excluded region is shaded in green in the left panel of \cref{fig:halomassf}. By confining the interaction to a narrow redshift range, the BP2 model satisfies this constraint while maintaining a statistical preference in CMB and WL data with $\chi^2-\chi^2_{\Lambda_{CDM}} = -11$.

However, the shape of the transfer function in our model differs from that of WDM, featuring stronger suppression at intermediate scales ($k \sim 1~h/\rm{Mpc}$) but less deviation from $\Lambda$CDM at smaller scales ($k \gtrsim 10~h/\mathrm{Mpc}$). To account for this, we calculate the subhalo mass function using the extended Press–Schechter formalism~\cite{Press:1973iz,Bond:1990iw,Esteban:2023xpk}. First, we compute the density field variance,
\begin{equation}
    \sigma^2(R(M))=\int\frac{d^3k}{(2\pi)^3}P_{\rm{lin}}(k)|W(k,R)|^2,
\label{sigma2}
\end{equation}
where $R(M)=3M/(4\pi\bar{\rho}_m)^{1/3}$ is the Lagrangian radius of a halo with mass $M$, $\bar{\rho}_m$ is the average matter density, $P_\textrm{lin}$ is the linear matter power spectrum, and $W(k,R)=(3/(kR)^3)\left[ \sin(kR) - kR \cos(kR) \right]$ is the window function. The halo mass function is then,
\begin{equation}
   \frac{dn_{halo}}{dM}=f(\sigma)\frac{\bar{\rho}_m}{M}\left|\frac{d\log \sigma(R(M) }{dM}\right|.
\label{halofunction}
\end{equation}
We used the Sheth-Tormen mass function~\cite{Sheth:2001dp}, $f(\sigma)=A[1+(\alpha\nu)^{-p}]\sqrt{a}e^{a\nu/2}/\sqrt{2\pi\nu}$, where $\nu=\delta_c/\sigma$ with the linear overdensity threshold for spherical collapse $\delta_c=1.686$, and we employ parameters $A=0.30$, $a=0.79$, and $p=0.22$~\cite{Esteban:2023xpk,Benson_2019}.

As shown in the right panel of \cref{fig:halomassf}, while halo formation in the BP2 model is strongly suppressed at masses around $M\sim 10^{11} - 10^{12}~\odot$, the predicted abundance of halos with mass $M\sim \textrm{a few}\times 10^8~M_\odot$ exceeds that in the WDM scenario, leading to good agreement with observations~\cite{DES:2020fxi}. High-resolution zoom-in simulations, which we leave for future work, are needed to verify this approximation.

Other benchmarks with a narrow interaction window, such as BP3 and BP4, worsen the fit to CMB and WL data, cf. \cref{tab:benchmarks}. For instance, BP4, which features a rapid interaction onset around matter-radiation equality (MRE), shows only a $1.1\sigma$ preference over $\Lambda$CDM and requires a fine-tuned cosmological impact of $\nu$DM interactions. Since the posterior distribution favors earlier interactions, we focus on the redshift range $z\sim 10^4-10^5$ in the following.

Recent studies have shown that $\nu$DM interactions with a constant $u_{\nu\textrm{DM}}\sim 10^{-4}$ are not only favored by CMB, BAO, and cosmic shear data but can also resolve the $S_8$ discrepancy~\cite{Zu:2025lrk}. As seen in \cref{tab:benchmarks}, our analysis yields $S_8$ values around $0.8$. This is slightly higher than in the ``$u_{\nu\textrm{DM}} = \textrm{constant}$'' scenario because the parameter 
$z_{\textrm{min}}$ shifts power suppression to smaller scales, affecting $\sigma_8$ less. This intermediate $S_8$ value aligns well with the recent Kilo-Degree Survey (KiDS) legacy analysis, which found $S_8=0.815^{+0.016}_{-0.021}$~\cite{Wright:2025xka}; see also the current status review in~\cite{CosmoVerse:2025txj}.

\section{Neutrino portal dark matter\label{sec:bsm}}

To further illustrate the impact of enhancing the $\nu$DM interaction rate within a limited redshift range, we consider a specific BSM interaction model where a DM fermion $\chi$ couples to a massive Dirac sterile neutrino $N$ via a complex scalar $\phi$ mediator, which carries a lepton number~\cite{Bertoni:2014mva,Batell:2017rol,Batell:2017cmf,Blennow:2019fhy}:
\begin{equation}
\mathcal{L} \supset -\lambda_\ell\,(\bar{L}_\ell\,\hat{H})\,N_R\ -\,\phi\,\bar{\chi}\,(y_L\,N_L+y_R\,N_R) + \textrm{h.c.}
\end{equation}
After electroweak symmetry breaking, this leads to $\chi$-$\nu$ couplings with a strength dependent on $y_L$ and $\nu$-$N$ mixing angles, $U_{\ell 4} \simeq \lambda_\ell v/m_N$, where $\ell = e, \mu, \tau$ and $m_N$ is the sterile neutrino mass. We take $U_{\tau 4}\neq 0$ and $U_{e 4} = U_{\mu 4} = 0$~\cite{Beacham:2019nyx}. We assume that $\chi$ is stabilized by a dark $U(1)_d$ symmetry and that $\Omega_\chi h^2 = 0.12$.\footnote{Adjusting the thermal DM relic density to the observed value~\cite{Planck:2018vyg} might require considering an asymmetric DM scenario or a non-standard early cosmological history. However, these considerations do not impact our results, which are determined by $\nu$DM interactions at later times, as long as the DM velocity distribution is not significantly affected.} In contrast, $\phi$ is unstable and decays via $\phi\to\chi\nu$ at $z\gg 10^5$. 

The $\nu$DM scattering in this model proceeds via the exchange of a $\phi$ mediator. The $s$-channel contribution yields a resonance at an incident neutrino energy $E_{\nu,\textrm{res}}\simeq (m_\phi^2-m_\chi^2)/2 m_\chi$. Assuming a small mass splitting $\delta = (m_\phi - m_\chi)/m_\chi \ll 1$, the relevant cross section plateaus in the energy range $E_{\nu,\textrm{res}} \lesssim E_\nu \lesssim m_\chi/2$. As $E_\nu$ approaches $E_\nu^{\textrm{res}}$ from above, $E_\nu \searrow E_{\nu,\textrm{res}}$, the cross section first increases sharply to its resonant value, $\sigma_{\textrm{res}}\simeq (g^4/16\pi)\,(\delta^2/\Gamma_\phi^2)$. Here, $\Gamma_\phi = (g^2/16\pi)\,m_\phi\,(1-m_\chi^2/m_\phi^2)^2$ is the $\phi$ decay width, and $g = y_L\sqrt{\sum_{\ell}{|U_{\ell r}|^2}}$. After peaking, the cross section becomes strongly suppressed at lower energies, $E_\nu\lesssim E_{\nu,\textrm{res}}$. In the high-energy regime, $E_\nu\gtrsim m_\chi/2$, the cross section also decreases with increasing energy, as $\sigma_{\nu\textrm{DM}}\propto 1/E_\nu$. Hence, $\sigma_{\nu\textrm{DM}}(E_\nu)$ is maximized in an energy regime around the resonance, therefore avoiding bounds relevant for both higher and lower $E_\nu$. We illustrate this in \cref{fig:resonantunuDM} with a brown solid line. For this illustration, we assumed a fixed neutrino energy at a given redshift, given by $E_\nu^2\simeq \langle E_\nu^2\rangle = 15\,[\xi(5)/\xi(3)]T_\nu^2 \simeq 12.94\,T_\nu^2$, obtained for the Fermi-Dirac distribution. 

Convolution with the actual redshift-dependent neutrino energy distribution significantly broadens this sharp peak. To study this broadening, we employ the ETHOS formalism to calculate the low-energy $\nu$DM elastic scattering rate in \cref{eq:thetachi}~\cite{Cyr-Racine:2015ihg,Stadler:2019dii},
\begin{equation}
\dot{\mu}_\chi = \frac{a\,n_\chi}{128\,\pi^3\,m_\chi^2\,(\rho_\nu/\eta_\nu)}\,\int_0^\infty{dp\,p^4\,\frac{\partial f^{(0)}_\nu}{\partial p}}\,A_{0-1}(p),
\label{eq:ETHOS}
\end{equation}
where $p$ is the incident neutrino momentum, we consider massless neutrinos with their unperturbed phase-space distribution function $f^{(0)}_\nu  = \eta_\nu/(\exp{(p/T_\nu)}+1)$, and the quantity $A_{0-1}(p) = A_0(p)-A_1(p)$ depends on the invariant matrix element of the $\nu$DM scattering process. For our model of interest, we find
\begin{equation}
A_{0-1}(p) = \frac{g^4m_\chi p^2\,[m_\chi+(8/3)p]}{[m_\chi^2+2m_\chi p-m_\phi^2]^2+m_\phi^2\Gamma_\phi^2},
\label{eq:A01}
\end{equation}
where we have focused solely on the dominant $s$-channel interaction. Comparing this to the case with a constant cross section, where $\dot{\mu}_\chi = a\,n_\chi\,\sigma_{\nu\textrm{DM}}$, we determine an effective $\nu$DM interaction cross section from \cref{eq:ETHOS} as $\sigma_{\textrm{eff},\nu\textrm{DM}} = \dot{\mu}_\chi/(a\,n_\chi)$. Subsequently, we define corresponding effective $u_{\textrm{eff},\nu\textrm{DM}}$ using the relation in \cref{eq:unuDM}. This parameter, as a function of redshift $z$, is shown by a red dashed line in \cref{fig:resonantunuDM} assuming $m_\chi = 1~\textrm{GeV}$, $\delta = 7\times 10^{-8}$, and $g = 2.3\times 10^{-4}$.\footnote{We note that the brown solid line in \cref{fig:resonantunuDM}, which corresponds to a fixed $E_\nu$, is shown assuming $g = 2.3\times 10^{-2}$ to better illustrate that the cross section plateaus at $E_\nu > E_{\nu, \textrm{res}}$. Otherwise, i.e. for $g = 2.3\times 10^{-4}$, the plateau would correspond to $u_{\nu\textrm{DM}}\ll 10^{-12}$. This change, however, does not affect the position of the $u_{\nu\textrm{DM}}$ peak as a function of $z$ and is therefore introduced purely for clarity.}

The low-energy tail of the neutrino's Fermi-Dirac distribution probes the $\nu$DM resonance even at high temperatures, $T_\nu\gg E_\nu^{\textrm{res}}$, which enhances the interaction rate over a range of redshifts compared to naive estimates. To assess the cosmological impact of this enhancement, we model the effective coupling, $u_{\textrm{eff},\nu\textrm{DM}}$, as a $100$-step piecewise function of redshift. Our benchmark scenario shows a mild improvement in fitting CMB and WL data of $\chi^2 - \chi^2_{\Lambda\textrm{CDM}}\simeq -12$ while satisfying numerous constraints. These include astrophysical bounds on $\nu$DM interactions, near-future direct or indirect DM detection searches~\cite{ Arguelles:2019ouk}, rare tau lepton decays~\cite{Batell:2017cmf}, and searches for heavy neutral leptons $N$. The last constraints are weakened if $m_N > m_\chi+m_\phi$, and $N\to\chi\phi^\ast$ is the dominant invisible decay channel.

However, evading small-scale structure constraints from dwarf galaxies requires suppressing the $\nu$DM scattering rate at high redshifts, $z\gtrsim \textrm{a few}\times 10^5$, a requirement illustrated by the MW satellite bound in \cref{fig:resonantunuDM}~\cite{Crumrine:2024sdn}. We achieve this suppression by introducing a minuscule time variation in the dark matter mass of order $10^{-8}\,m_\chi$. While such a small variation is undetectable in traditional DM searches, it strongly impacts the near-resonant $\nu$DM interactions, providing the necessary effect; see~\cite{Anderson:1997un,Das:2006ht,Davoudiasl:2019xeb,Boubekeur:2023fqo,Das:2023enn,Chakraborty:2024pxy} for discussions of the cosmological impact of the time-varying DM mass.

We model the DM mass variation by its coupling to a background ultralight scalar field $\varphi$ with a quadratic potential, $V(\varphi) = \frac{1}{2}m_\varphi^2\varphi^2$. This yields the field-dependent DM mass,
\begin{equation}
m_\chi(\varphi) = m_\chi(0)\,(1+g_\varphi\,\varphi^2/2),
\end{equation}
where $g_\phi$ is the coupling. The field is initially damped but begins to oscillate when $3H(t)\sim m_\varphi$. Assuming a zero initial derivative, the field's evolution during the radiation-dominated epoch is described by~\cite{OHare:2024nmr}
\begin{equation}
\varphi(t) = \phi_i\,\left(\frac{2}{m_\varphi t}\right)^{1/4}\,\Gamma(5/4)\,J_{\frac{1}{4}}(m_\varphi t),
\end{equation}
where $J_{1/4}$ is the Bessel function.

We assume the two dark sector masses are nearly degenerate at high redshifts ($\delta\lesssim 10^{-9}$), with the mass splitting between them growing as the background field starts to oscillate. To minimize the fitted $\chi^2$ for this illustration, we use a benchmark with $g = 1.5\times 10^{-4}$, $m_\varphi = 2\times 10^{-24}~\textrm{eV}$, and $g_\varphi \varphi_i^2 = 1.4\times 10^{-7}$.\footnote{The oscillating ultralight scalar field additionally contributes to the DM density in the universe. However, this contribution remains strongly suppressed and subdominant for the model parameters of our interest~\cite{OHare:2024nmr}.} The tiny mass splitting suppresses the $\nu$DM scattering rate at high redshifts, where the resonant regime corresponds to the very low-energy tail of the neutrino energy distribution. At later times, the resonant enhancement becomes dominant, thereby increasing $u_{\textrm{eff},\nu\textrm{DM}}$ by a few orders of magnitude, as shown by the blue dot-dashed line in \cref{fig:resonantunuDM}.

This benchmark model allows for a fit to CMB and WL data with  $\chi^2 - \chi^2_{\Lambda\textrm{CDM}} \simeq -5$, while avoiding excessive suppression of small-scale structures. For comparison, \cref{fig:halomassf} shows its transfer function and the subhalo mass function. For masses $M\lesssim \textrm{a few}\times 10^8\,M_\odot$, its predictions are closer to $\Lambda$CDM than those of the sample WDM model constrained by the MW satellite data~\cite{DES:2020fxi}.

Finally, in the ``$u_{\nu\textrm{DM}} = \textrm{const}$'' scenario, the Lyman-$\alpha$ forest data alone yield a best-fit $\nu$DM interaction strength for massless neutrinos that is lower than preferred by the CMB and WL data, $u_{\nu\textrm{DM}, \textrm{Lyman-}\alpha}\simeq 3.8\times 10^{-6}$~\cite{Hooper:2021rjc}. This discrepancy, potentially due to modeling issues (e.g., baryonic feedback), could also be explained in the $\nu$DM portal model. Notably, the Lyman-$\alpha$ forest probes smaller physical scales ($k\gtrsim 5~h/\rm{Mpc}$) than WL. The suppression of the matter power spectrum observed at these scales could reflect the predicted redshift-dependent decrease of $u_{\nu\textrm{DM}}$. As shown in the left panel of \cref{fig:halomassf}, our benchmark model produces a matter power spectrum suppression at $k\sim 5~h/\rm{Mpc}$ that is similar to the Lyman-$\alpha$ best-fit, although the two predictions diverge at smaller scales where stringent dwarf galaxy bounds apply.

\section{Conclusions\label{sec:conclusions}}

Neutrinos and dark matter, predicted as undetected early-universe relics, could have significantly shaped cosmological perturbations through their interactions. Recent analyses of Lyman-$\alpha$ forest, CMB, and weak lensing data suggest non-zero $\nu$DM interactions. However, in the simplest scenario of a constant $\nu$DM cross section, the best-fit value conflicts with other cosmological and astrophysical probes.

This approach, however, generally fails to capture the non-trivial temperature dependence of the $\nu$DM interaction characteristic of realistic BSM scenarios. Specifically, the $\nu$DM interaction rate can be maximized within a particular redshift range, imprinting distinct signatures on large-scale structures at various scales. We have demonstrated that such an enhancement of the $\nu$DM interaction strength at redshifts $z\sim (10^4-10^5)$ reproduces the previously observed preference for non-zero $u_{\nu\textrm{DM}}$ in CMB and cosmic shear data. Crucially, this mitigates constraints from astrophysics and small-scale structures. We illustrated this using a specific neutrino-DM portal scenario that utilizes a resonantly-enhanced $\nu$DM scattering cross section convoluted with a broad neutrino energy spectrum, which varies in the expanding universe. This effect can be further influenced by even minuscule time variations of the DM mass.

While narrow-width features in DM interactions are typically overlooked in cosmological modeling, they can impact a wide range of perturbations when they are sensitive to radiation components prior to matter-radiation equality. Resonant scatterings have long served as a fundamental discovery tool in accelerator experiments. Their cosmological impact deserves similar attention.

\acknowledgments

We thank Eleonora Di Valentino, William Giarè, and Yue-Lin Sming Tsai for useful comments on the manuscript. This work is supported by the National Science Centre, Poland (research grant No. 2021/42/E/ST2/00031). LZ is also supported by the NAWA Ulam fellowship (No. BPN/ULM/2023/1/00107/U/00001).


\appendix
\section{Details of cosmological data analysis\label{app:simulation}}

In this work, we use the following cosmological data in the likelihood analysis:

\begin{enumerate}
    \item[(i)] DES Three-year cosmic shear observations~\cite{DES:2021bvc}. We refer to this dataset as DES Y3 throughout this work. The calculation of the cosmic shear likelihood requires the nonlinear matter power spectrum. Following~\cite{Zhang:2024mmg,Zu:2025lrk}, we employ a well-developed emulator, which is based on $200$ $N$-body simulations with dark acoustic oscillation initial conditions, to perform the nonlinear corrections. We conservatively mask scales beyond $k \sim \textrm{a few}~h/\rm{Mpc}$. Future weak lensing surveys, with robust modeling of smaller scales, could improve $\nu$DM bounds in the high-redshift regime~\cite{S:2024xgp}. 
    
    \item[(ii)] The CMB likelihoods from Planck 2018 Legacy~\cite{Planck:2019nip}, including high-$\ell$ power spectra (TT, TE, and EE) cut at $l<650$, low-$\ell$ power spectra (TT and EE), and the Planck lensing reconstruction. We refer to this dataset as Planck.
    \item[(iii)] The full ACT temperature and polarization DR4 likelihood~\cite{ACT:2020frw}.
    \item[(iv)] The BAO likelihood with measurements from the BOSS DR16~\cite{eBOSS:2020tmo,eBOSS:2020hur,eBOSS:2020uxp,eBOSS:2020fvk,eBOSS:2020yzd}.
\end{enumerate}

In our analysis, we adopt the implementation of $\nu$DM interactions within the \texttt{CLASS} code, as detailed in~\cite{Stadler:2019dii,Mosbech:2020ahp}. To model the resonantly-enhanced $\nu$DM interactions in the neutrino portal model, we calculate the effective $u_{\textrm{eff},\nu\textrm{DM}}$ parameter as outlined in the text. For numerical computation, we implement a discretized approximation of the $u_{\textrm{eff},\nu\textrm{DM}}(z)$ function. We have verified for selected benchmark scenarios that employing $n_{\textrm{step}} \geq 100$ steps yields convergent results. Therefore, we use the $100$-step implementation in our analysis.

\begin{table}[!ht]
    \centering
    \begin{tabular}{|c|c|c|c|c|c|c|c|}
\hline
\hline
$100 \Omega_b h^2$  & [2.147, 2.327] \\
$\Omega_{\rm{CDM}} h^2$ &   [0 , 0.2] \\
$100 \theta_s$ &  [1.0393,  1.0429] \\
$\ln \left(10^{10} A_s\right)$  & [2.9547,  3.1347] \\
$n_s$ &   [0.9407, 0.9911]  \\
$\tau_{\text {reio }}$ &    [0.01,  0.7]  \\
$\log_{10}\rm{u}_{\nu\rm{DM}}$  & [-8.0, 0.0] \\
$\log_{10}z_{min}$  & [1, 6]\\
$\log_{10}\Delta z$  & [1, 6]\\
\hline
\hline
\end{tabular}
    \caption{The parameters and their prior ranges used in the MCMC analysis are listed above. We use flat priors within the specified ranges.}
    \label{tab:prior}
\end{table}

\begin{figure*}[t!]
\centering
\includegraphics[width=1\textwidth]{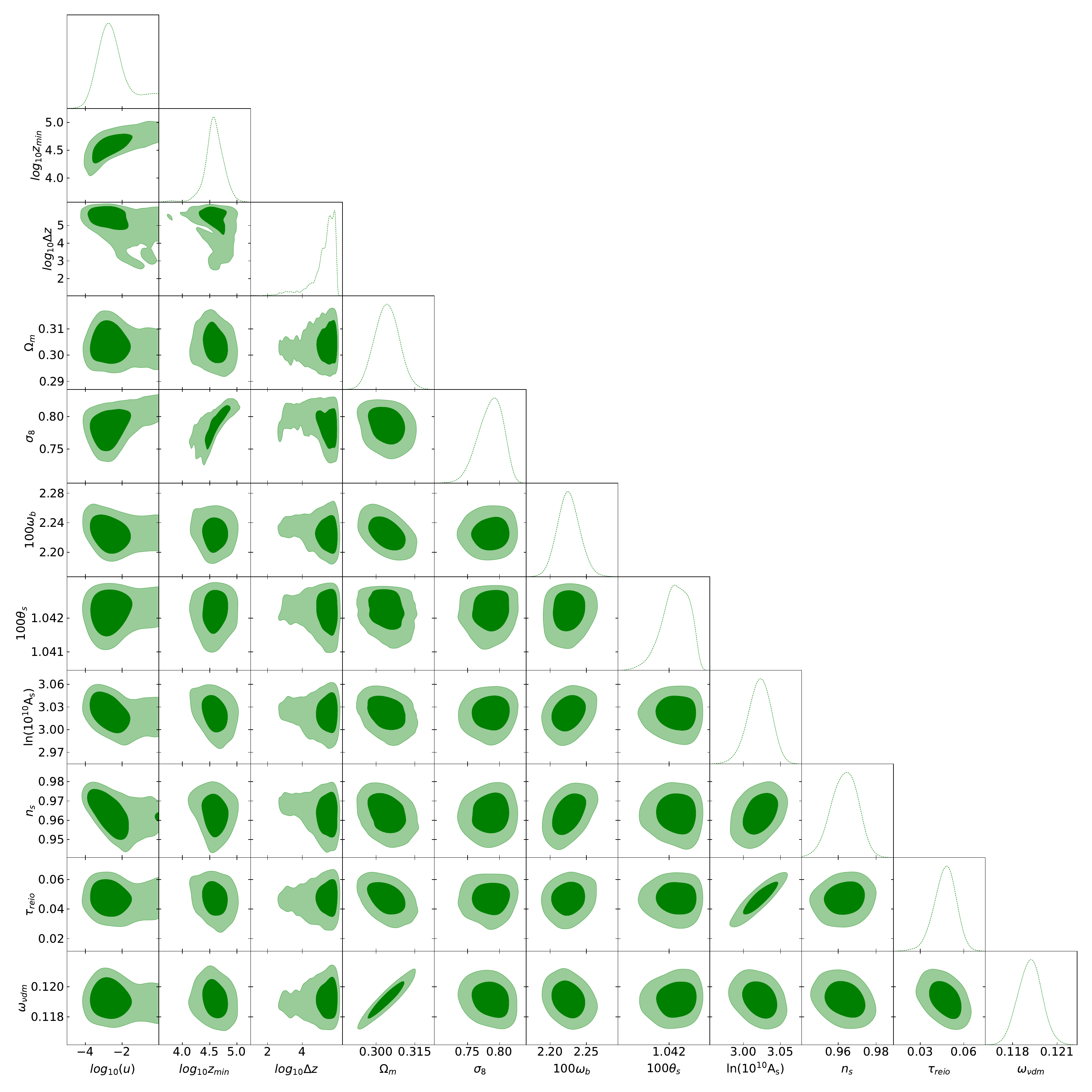}
\caption{The triangle plot for the MCMC scan}
\label{fig:triangle}
\end{figure*}

We perform MCMC parameter scans using the \texttt{MontePython} code~\cite{Audren:2012wb}. The parameter prior ranges are presented in \cref{tab:prior}. The triangle plot obtained in the analysis presented is shown in \cref{fig:triangle}. 

It is worth noting that the one-dimensional posterior distribution function for $\textrm{u}_{\nu\textrm{DM}}$ exhibits a long tail, indicating a preference for non-zero values at the $2\sigma$ level. This tail is partially driven by models where the $\nu$DM impact is increasingly stronger (larger $\textrm{u}_{\nu\textrm{DM}}$) but better localized in time around $z \sim 10^5$. Specifically, the observed tail corresponds to a lower limit of $\textrm{u}_{\nu\textrm{DM}} > -3.65$. If $\nu$DM interactions remain effectively non-negligible only within a specific redshift range, the cosmological data favor a non-zero interaction strength that is larger than in the scenario where $\textrm{u}_{\nu\textrm{DM}}$ is constant.

\bibliographystyle{apsrev4-1}
\bibliography{biblio}{}
\end{document}